\documentclass[12pt]{iopart}

\usepackage{graphicx}
\usepackage{dcolumn}
\usepackage{bm}
\usepackage{ulem}

\newcommand{\ket}[1]{\mbox{$ | #1 \rangle $}}

\begin{document}

\title[Waveguide polarization entanglement source at a telecom wavelength]{Polarization entangled photon-pair source based on a type-II PPLN waveguide emitting at a telecom wavelength}

\author{A. Martin$^1$, A. Issautier$^1$, H. Herrmann$^2$, W. Sohler$^2$, D. B. Ostrowsky$^1$, O. Alibart$^1$, and S. Tanzilli$^1$}%
\address{$^1$ Laboratoire de Physique de la Mati\`ere Condens\'ee, CNRS UMR 6622,
Universit\'e de Nice~-~Sophia Antipolis, Parc Valrose, 06108 Nice Cedex 2, France.}
\address{$^2$ Angewandte Physik, Universitat-GH-Paderborn, Postfach 1621, D-4790 Paderborn, Germany}
\ead{sebastien.tanzilli@unice.fr}

\begin{abstract}
We report the realization of a fiber coupled polarization entangled photon-pair source at 1310\,nm based on a birefringent titanium in-diffused waveguide integrated on periodically poled lithium niobate. By taking advantage of a dedicated and high-performance setup, we characterized the quantum properties of the pairs by measuring two-photon interference in both Hong-Ou-Mandel and standard Bell inequality configurations. We obtained, for the two sets of measurements, interference net visibilities reaching nearly 100\%, which represent important and competitive results compared to similar waveguide-based configurations already reported. These results prove the relevance of our approach as an enabling technology for long-distance quantum communication.
\end{abstract}

                            
\maketitle

\section{\label{sec:level1}Introduction}

Quantum superpositions of states and entanglement are now considered to be resources for quantum communication protocols such as quantum key distribution~\cite{gisin_quantum_2002}, quantum teleportation~\cite{kim_quantum_2001}, entanglement swapping~\cite{halder_entangling_2007}, quantum relays~\cite{collins_quantum_2005}, and quantum repeaters~\cite{briegel_1998,duan_2001,simon_2007}. Quantum bits of information (qubits) can be encoded on photons using different properties, the most convenient being polarization and time-bin~\cite{weihs_photonic_2001}. Experimentally, spontaneous parametric down-conversion (SPDC) in non-linear bulk crystals~\cite{kwiat_ultrabright_1999,fedrizzi_wavelength-tunable_2007,fiorentino_compact_2008,hentschel_2009}, waveguide crystals~\cite{tanzilli_ppln_2002,halder_high_2008}, and optical fibers~\cite{lee_generation_2006,medic_2010}, has up to now been widely employed to produce polarization~\cite{kwiat_ultrabright_1999,fedrizzi_wavelength-tunable_2007,fiorentino_compact_2008,lee_generation_2006,medic_2010} or time-bin~\cite{tanzilli_ppln_2002,halder_high_2008} entangled photon-pairs either in the visible~\cite{kwiat_ultrabright_1999,fedrizzi_wavelength-tunable_2007,fiorentino_compact_2008} or telecom~\cite{hentschel_2009,tanzilli_ppln_2002,halder_high_2008} range of wavelength.

However, with the emergence of ``out of the lab'' quantum communication links over long distances~\cite{lloyd_long_2001,landry_quantum_2007,Ursin_entanglement_2007,hubel_2007}, there is a need for new generation sources featuring higher efficiencies, better stability, compactness, and, most of all, showing a near perfect quality of entanglement. In this case, it is also necessary to produce entanglement carried by photons compatible with telecommunications fibers that show transparency windows centered at 1310 and 1550\,nm. In addition, photons have to be generated over narrow bandwidths so as to avoid both chromatic and polarization mode dispersions (PMD) in these fibers~\cite{fasel_quantum_2004}, while preventing from multiple-pair production in the bandwidth and time windows of interest~\cite{marcikic_2002,de_riedmatten_2004}. Note also that because of both PMD and time-varying birefringence in the fiber links, reliable distribution of polarization qubits can be an issue. Several schemes have been proposed to either actively or passively compensate for such undesirable effects leading to quantum state decoherence~\cite{hubel_2007,ursin_danube_2004,Peng_2007,Xavier_2008,lucio-martinez_2009}.

In this framework, integrated and fiber non-linear optics appear to be natural and very promising candidates, offering the possibility to efficiently create polarization entanglement associated with a high compatibility with standard fiber networks. However, most of the previous realizations involving integrated non-linear waveguides led to limited quantum properties when two-photon quantum interference was measured either using a Hong-Ou-Mandel (HOM) setup~\cite{hong_measurement_1987,fujii_bright_2007,martin_integrated_2009,Caillet_dip_contra_2009} or a Bell inequality-type setup~\cite{Bell_1964,clauser_proposed_1969,suhara_generation_2007,takesue_generation_2005,jiang_generation_2007,kawashima_type_I_2009}. Note that Ref.~\cite{zhong_high_2009} reported a HOM setup leading to a net visibility of 98.5\%, but no entanglement was demonstrated. 
Regarding fiber sources, very high-quality entanglement has been reported~\cite{lee_generation_2006,medic_2010}. 

We show in this paper that our solution based on a type-II, titanium (Ti) in-diffused periodically poled lithium niobate (PPLN) waveguide associated with a 50/50 beam-splitter (BS) offers a simple and controllable way to achieve polarization entanglement at a telecom wavelength. After a brief outline of the state of the art, we describe our source setup and detail the key points for which particular attention has been paid to enable reaching optimum entanglement. Then, we depict how we analyze the quantum properties of our photon-pairs. By the use of an original single configuration, we can perform two complementary quantum tests. On the one hand, the photons indistinguishability is determined and accurately adjusted using a HOM-type experiment in order to maximize the potential amount  of entanglement~\cite{hong_measurement_1987}. On the other hand, entanglement is analyzed using a standard Bell inequality-type measurement~\cite{Bell_1964,clauser_proposed_1969}. For both types of tests, we achieved near-perfect two-photon interference visibilities, competitive with any other reported scheme, and proving the relevance of our overall approach: using guided-wave optics as an enabling technology for quantum information generation and manipulation.

\section{Guided-wave optics for polarization entangled photons emitted at a telecom wavelength: state of the art}

There have been various demonstrations of polarization entanglement sources based on non-linear guided-wave optics, see \tablename{~\ref{tab:inv}} for a comparison. Takesue \textit{et al.} proposed a realization in which time-bin entanglement at 1550\,nm was converted to polarization entanglement by the use of an orthogonal polarization delay circuit~\cite{takesue_generation_2005}. In this configuration however, the quality of entanglement was limited by the stability of this delay circuit, mainly due to the arms relative phase fluctuations. They obtained a net visibility of 82\% for entanglement measurements.
To avoid the phase stabilization issue, Jiang \textit{et al.} proposed a Sagnac interferometric approach in which two separate type-I PPLN waveguides were mounted in a fiber loop to generate paired photons at the non-degenerate wavelengths of 1434 and 1606\,nm~\cite{jiang_generation_2007}. With such a configuration, they obtained a raw visibility of 93.5\%. Based on a comparable Sagnac setup surrounding one type-0 PPLN waveguide, Lim \textit{et al.} demonstrated a net fidelity exceeding 96\% to the $\ket{\Phi^+}$ entangled state, for photons emitted at 1542 and 1562\,nm~\cite{lim_2008}.
Moreover, Kawashima \textit{et al.} proposed an original solution in which they employed a single type-I Ti in-diffused PPLN waveguide, having a half-wave plate (HWP) inserted in the middle along its propagation axis~\cite{kawashima_type_I_2009}. They obtained an entangled state leading to a net visibility of 83\%.

There have been many sources based on non-linear $\chi^{(3)}$ fibers as well, either relying on dispersion-shifted fiber (DSF) or photonic crystal fiber (PCF) technologies, but only a few emitting at a telecom wavelength~\cite{lee_generation_2006,medic_2010,mcmillan_2009}.
One of the most relevant and recent realizations produces telecom entangled photons via bichromatic pulses pumping a DSF placed in a Sagnac loop aligned to deterministically separate degenerate photon pairs~\cite{medic_2010}. The authors obtained a state fidelity to the nearest maximally polarization entangled state of 0.997 when subtracting background Raman photons, multiple pair generation events, and accidental coincidences initiated by detectors dark counts. In this experiment, note that the Raman effect was globally reduced by cooling down the fiber to the temperature of 77\,K (LN$_2$). 

In all of the above mentioned experiments, the created entangled states were of the form $\ket{\Phi^\pm} = \frac{1}{\sqrt{2}} \left[ \ket{H_a,H_b} \pm \ket{V_a,V_b}\right]$, where the indices $a$ and $b$ label the two photons, and $H$ and $V$ represent horizontal and vertical polarization modes, respectively. We will now see how we can implement a simple solution based on a single type-II Ti in-diffused PPLN waveguide and a 50/50 BS, while avoiding issues associated with either an interferometric setup or a polluting spontaneous emission. Using the same notations as before, the generated state is in this case of the form $\ket{\Psi^\pm} = \frac{1}{\sqrt{2}} \left[ \ket{H_a,V_b} \pm \ket{V_a,H_b}\right]$.

\section{\label{sec:level2}Setup of the source}

\begin{figure}[t]
\resizebox{1\columnwidth}{!}{\includegraphics{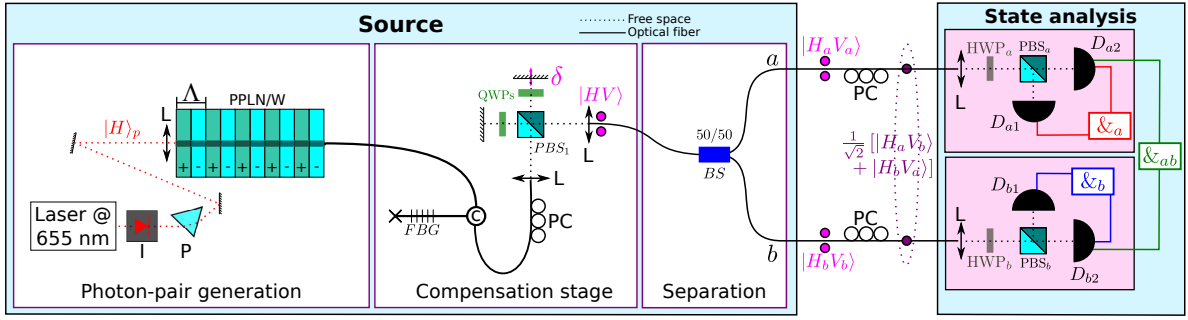}}
\caption{\label{fig:montage}Experimental setup of the source and analysis system. The pairs of circles represent pairs of photons with their associated polarization quantum states, after the birefringence compensation system and after the beam-splitter (output of the source). P: prism used to filter any infrared light out of the laser; I: optical isolator; L: lenses; PPLN/W: periodically poled lithium niobate waveguide; FBG: tunable fiber Bragg grating filter centered at 1310\,nm with a bandpass of 0.5\,nm; C: fiber optical circulator; PC: polarization controller; PBS: polarization beam-splitter; QWP: quarter wave-plate; BS: 50/50 coupler; HWP: half wave-plate; SB: Soleil-Babinet compensator; D: single photon detectors; and $\&$: AND-gate that can be placed between any two detectors depending on the performed measurement.}
\end{figure}

The setup of the source is depicted in \figurename{~\ref{fig:montage}}. Photon-pair generation is obtained by SPDC in a type-II Ti in-diffused PPLN waveguide. In this case, H-polarized photons from a pump field ($p$) can be converted into two cross-polarized photons usually called signal ($s$) and idler ($i$), \textit{i.e.}, in the product state $\ket{H_s,V_i}$. SPDC process is ruled by the conservation of both energy, $\omega_p = \omega_s + \omega_i$, and momentum, $\overrightarrow{k}_p = \overrightarrow{k}_s + \overrightarrow{k}_i$. The latter equation is also known as the phase matching condition which, in the case of periodically poled crystals, becomes the so-called quasi phase matching (QPM), $\overrightarrow{k}_p = \overrightarrow{k}_s + \overrightarrow{k}_i + \frac{2\pi}{\Lambda}\overrightarrow{u}$, where $\Lambda$ and $\overrightarrow{u}$ represent the poling period and a unit vector perpendicular to the domain grating, respectively. Compared to bulk realizations, the use of an integrated non-linear $\chi^{(2)}$ waveguide permits enhancing SPDC thanks to the high confinement of the three interacting waves over a small transverse area, on the order of tens of $\mu$m$^2$, and over several centimeters of length.

Starting with a CW laser at 655\,nm, the idea is to generate paired photons at the degenerate wavelength of 1310\,nm, corresponding to the second telecom window. We have calculated, simulated, and verified experimentally that the required poling period is 6.6\,$\mu$m when associated with a waveguide width of 5\,$\mu$m and an operation temperature of 96.8\,$^\circ$C. More details regarding our numerical simulations, fabrication processes, and classical characterizations are given in Ref.~\cite{martin_integrated_2009}.

\begin{figure}[t]
\begin{center}
\resizebox{0.6\columnwidth}{!}{\includegraphics{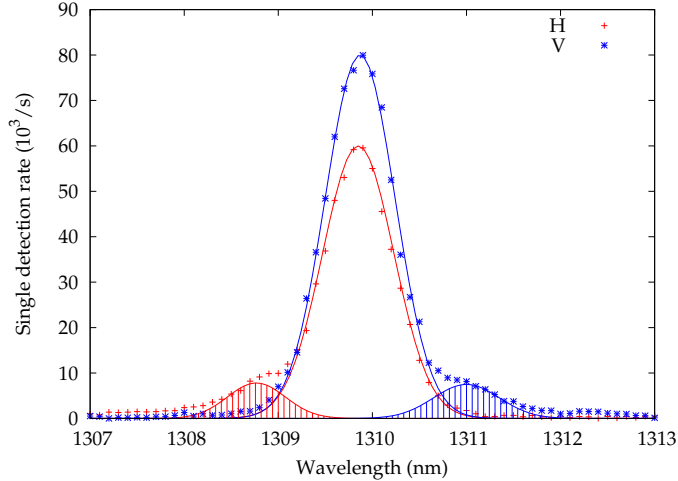}}
\caption{\label{fig:spectrum}
Emission spectra out of the waveguide as a function of the wavelength, for the two polarization modes. Here the laser wavelength
is fixed at 654.9\,nm. The observed spectrum could be fitted assuming two distinct QPM solutions (empty and filled curves): one leading to paired photons at the degenerate wavelength of interest, \textit{i.e.}, of 1309.8\,nm, for the pump in the fundamental mode, and the other with the
H photon at 1308.7\,nm and the V photon at 1310.9\,nm.}
\end{center}
\end{figure}

Obtaining a maximally entangled state of the form $\frac{1}{\sqrt{2}}\left[\ket{H_a,V_b}+\ket{V_a,H_b}\right]$ from two cross-polarized photons and a 50/50 BS is only possible provided the paired photons are made perfectly indistinguishable before the BS for all degrees of freedom, but the polarization~\cite{fujii_bright_2007,martin_integrated_2009,suhara_generation_2007,zhong_high_2009}.
Here, we are concerned by spectral, spatial, and temporal observables. This is far from being the case at the direct output of the waveguide. We will now outline various identified key points to which particular attention has to be provided for reaching high-quality interference for both indistinguishability and entanglement measurements. 

First, birefringence in the crystal gives rise to two transverse intensity profiles at 1310\,nm (spatial modes), each associated with one of the two polarization modes. To erase this spatial mismatch, we collect the photon-pairs in a standard single-mode fiber.

Second, this birefringence also leads to a time delay between the two created photons since $H$ and $V$ modes are associated with different group velocities. To compensate for this temporal distinguishability, we designed and added on the path of the paired photons an adjustable birefringence compensator which has the arrangement of a polarization interferometer.
This amounts to separating the paired photons regarding their polarization states with a polarization beam-splitter (PBS), and to adjusting the delay between the two arms with a motorized mirror. A quarter-wave plate (QWP) is inserted in each arm enabling, over the round trip of the corresponding single photon, a 90$^\circ$-rotation of its polarization state. This way, the two photons leave the interferometer together through the other output, avoiding additional losses except those due to fiber coupling~\cite{wolfgramm_bright_2008}. It is important to note that in this compensation system no interference occurs since the two paths are associated with orthogonal polarization modes. Moreover, any phase drift in this system will be transposed as a global phase factor in the entangled state further created at the 50/50 coupler. No device stabilization is therefore required in opposition to Ref.~\cite{takesue_generation_2005}.

Third, we studied the emission spectrum produced by the waveguide in the single photon counting regime. As we can see in \figurename{~\ref{fig:spectrum}}, the central peak shows a spectral width of 0.7\,nm as predicted by theory \cite{martin_integrated_2009}.
The vertically polarized emission shows a sideband on the long wavelength region, while the horizontal polarized shows a sideband on the short wavelength region. The spectral shape could be well fitted assuming two distinct QPM solutions.
This particular spectral response is probably due to the combination of inhomogeneities in the waveguide (Ti-layer thickness, width, and poling period) initiated during the fabrication, associated with the multi-mode character at the pump wavelength~\cite{zhong_high_2009,fiorentino_spontaneous_2007}.
To overcome this spectral issue, we used a tunable fiber Bragg grating (FBG) filter having a FWHM of 0.5\,nm, \textit{i.e.}, slightly smaller than the natural bandwidth of the generated photons (see \figurename{~\ref{fig:spectrum}}). This way, only the paired photons having wavelengths belonging to the central peak are selected and the other peaks discarded. As a matter of fact, we have estimated the sideband to be 15\% of the central peak which corresponds very well with the reduced visibility of 85\% reported in our previous paper~\cite{martin_integrated_2009}. This observation is a clear indication that perfect entanglement is not possible without the use of a narrow filter to remove these polluting sidebands.

After all the compensations are done, we separate the two photons at a 50/50 coupler to create the state $\frac{1}{\sqrt{2}}\left[\ket{H_a,V_b}+\ket{V_a,H_b}\right]$. This is obtained when the separation is successful, \textit{i.e.}, in half of the cases. Then, the entangled photons are distributed to Alice ($a$) and Bob ($b$), respectively.

\section{\label{sec:level3}Quantum characterization of the source}

As already mentioned, we can perform two different quantum tests with this setup, namely two-photon interference experiments using either a HOM configuration when the photons are not separated at the BS, or a Bell inequality-like configuration when they are separated.
Whatever the type of measurement we perform, photon-pair detection is ensured by a passively-quenched germanium avalanche photodiode (Ge-APD) that triggers an InGaAs-APD (id Quantique id201) operated in gated mode. These APDs feature 4\% and 10\% quantum efficiencies and dark count probabilities of about $2.2 \times 10^{-5}$/ns and $10^{-5}$/ns, respectively. A dedicated electronics enables recording both the single and the coincidence (AND-gate) counts in real time. Note that for all the following measurements, we obtained average single and coincidence rates of 85\,kcounts/s and 450\,counts/s, respectively, for an injected pump power of 2.5\,mW in the waveguide. In terms of efficiency, the waveguide shows a brightness corresponding to $3\cdot10^5$ pairs created per second, per GHz of bandwidth, and per mW of pump power at its direct output~\cite{martin_integrated_2009}. In the meantime, the losses in the setup were measured to be on the order of 10.5 dB from the waveguide to Alice and Bob's detectors~\cite{tanzilli_ppln_2002}, therefore allowing us to infer a mean number of photons per relevant time window of 0.097. In our configuration, the coincidence time window has been set to 1.5\,ns to accommodate the APDs' timing jitter which is much larger than both the photon's coherence time and chromatic dispersion in the employed fibers.

\begin{figure}[t]
\centering
\resizebox{0.6\columnwidth}{!}{\includegraphics{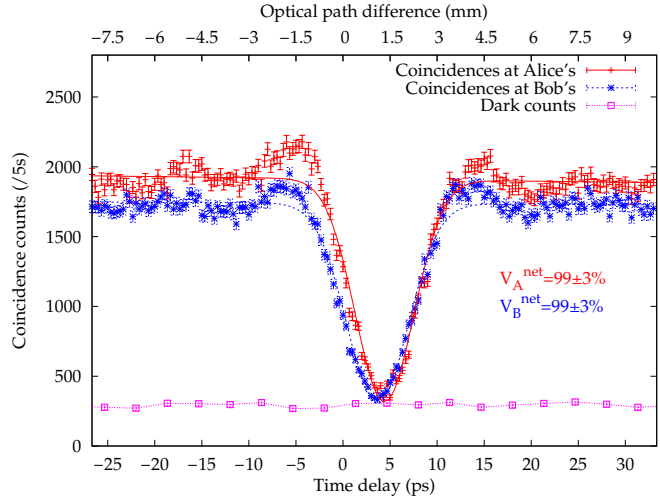}}
\caption{\label{fig:dip}Results for the HOM-type measurement obtained at the two users locations. The uncertainty associated with each point have been calculated using standard squared root deviation. We used a fitting function to estimate the visibilities and their associated asymptotic standard errors. We obtained net (raw) visibilities of $99\pm3\%$ ($83\pm2\%$). Here the corresponding subtracted noise, or accidental coincidence events, is only due to intrinsic dark counts in the employed APDs.}
\end{figure}

To perform the quantum measurements, Alice and Bob each have an adjustable polarization analyzer consisting of a HWP and a PBS in order to project the incoming state onto any linear polarization basis. In order to maximize the potential amount of entanglement, we first realized a HOM-like two-photon interference measurement taking advantage of the cases for which the photons are not separated at the BS. In these cases they arrive at Alice's or Bob's positions in the $\ket{H,V}$ state. This allows determining the photons indistinguishability in terms of spatial, temporal and spectral overlaps by analyzing the state $\ket{H,V}_j$ (j=a,b) in the $\{D,A\}$ basis for diagonal and anti-diagonal polarization, respectively. This is obtained when the corresponding HWP$_j$ in the analyzer is rotated by an angle of 22.5$^\circ$. In this basis, the state can be written as $\frac{1}{2} \left[\ket{D,D}-\ket{A,A} +\ket{D,A} -\ket{A,D} \right]$. If the two incoming photons are indistinguishable, $\ket{A,D}$ and $\ket{D,A}$ interfere destructively so that the state can be reduced to $\frac{1}{\sqrt{2}} \left[\ket{D,D} -\ket{A,A}\right] $, with correct normalization. In other words, in this basis two indistinguishable photons are projected onto a coherent superposition of states in which they always have the same polarization. As a consequence, they always exit through the same output port of the PBS. This leads to a HOM-like dip in the coincidence counts recorded between the considered pair of detectors (D$_{j1}$ and D$_{j2}$).
To observe such a dip experimentally, we adjust the temporal overlap with the moving arm of the birefringence compensation system, which makes the coincidence rate between D$_{j1}$ and D$_{j2}$ drop to zero when the overlap is maximum. By this measurement, we can infer the photons indistinguishability from the net visibility, $$V^{net} = \frac{R^{max,net}_c-R^{min,net}_c}{R^{max,net}_c}= \frac{R^{max}_c-R^{min}_c}{R^{max}_c -R^{acc}_c},$$ where $R^{max}_c$ ($R^{min}_c$) is the coincidence rate outside (inside) the dip, $R^{acc}_c$ represents the accidental coincidence rate only due to the dark counts in the detectors, and $R^{net}_c = R_c - R^{acc}_c$. In \figurename{~\ref{fig:dip}} we show the obtained dips at both locations each showing a net (raw) visibility of $99\pm3\%$ ($83\pm2\%$), comparable to the best results in any configuration reported to date. 

As described in Refs.~\cite{hong_measurement_1987,DeRiedmatten_Qint_SpatialSeparated_03}, the shapes of the dips are given by the convolution of the two wave-packets arriving at each PBS$_j$ placed in Alice's and Bob's analyzer, respectively. Due to the convolution product, the expected FWHMs of the dips are given by the relation $\tau_{dip,j} = \sqrt{2} \cdot \tau_{coh}$, where $\tau_{coh}$ the coherence time of the photons which can directly be obtained from their FWHM spectral bandwidth.
The FWHM of both dips are measured to be approximatively 7.45\,ps, leading to a coherence time of about 5.2\,ps. The latter result is in very good agreement with the theoretical value obtained from the FBG's FWHM bandwidth given by $\tau= 0.44 \frac{\lambda^2}{c\Delta\lambda}\simeq 5.0$\,ps, where $\Delta \lambda$ corresponds to the FBG bandwidth (0.5\,nm). 

Apart from estimating the quality of indistinguishability in the two arms, such a measurement also permits quantifying the difference of polarization mode dispersion (PMD), accumulated over the transmission fiber channels and through the various optical elements, by comparison of the relative position of the two dips. In our experiment, we can see from \figurename{~\ref{fig:dip}} that these dips are very close to one another since we employed fibers of only several tens of meter long.

Finally note that, in the case of actual quantum cryptography applications, Alice and Bob can independently, perform such a HOM measurement taking advantage of unseparated pairs, making it possible to check whether the paired photons are still indistinguishable in real time and therefore whether the source is still running well.

The latter measurements allowed determining the maximum degree of entanglement available at the output of the BS when the two photons are actually separated. Intuitively, this may correspond to the value of the overlap integral between the two obtained dips at both locations. One has therefore to choose the intermediate position for the moving mirror in the birefringence compensator to reach this maximum. 
Now that the photons are proven indistinguishable, we can assess the degree of entanglement using a Bell inequality-type interference experiment. This amounts to studying the visibility of the coincidence rate as a function of the relative angle between Alice and Bob's analyzers, in opposition to other known methods such as fidelity, tangle, or concurrence measurements~\cite{altepeter_entanglement_2005}. More precisely, two complementary settings are necessary for HWP$_a$, namely 0$^\circ$ ($\{H,V\}$ basis) and 22.5$^\circ$ ($\{D,A\}$ basis), and we observed the coincidence evolution as a function of the angle of HWP$_b$. The visibilities of these curves are related to the quality of both the optical setup ($\theta_{HWP_a} = 0^\circ$) and the entanglement ($\theta_{HWP_a} = 22.5^\circ$), respectively.

\begin{figure}[htb]
\centering
\resizebox{0.65\columnwidth}{!}{\includegraphics{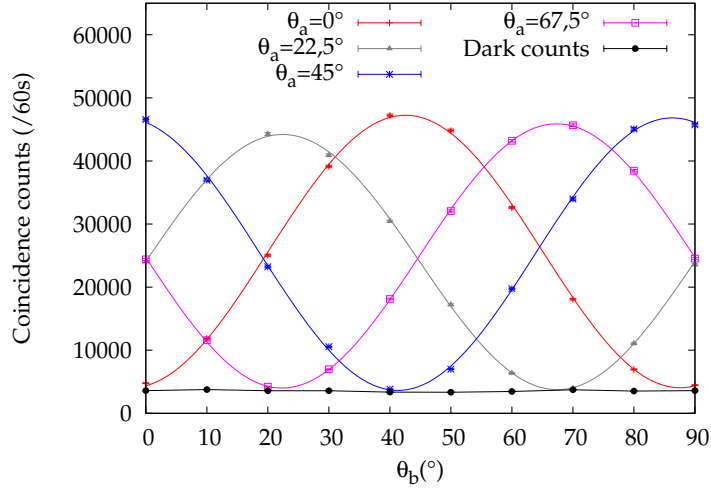}}
\caption{\label{fig:Bells}Obtained results for a standard entanglement measurement in the $\{H,V\}$ and $\{D,A\}$ basis. The uncertainty associated with each point is barely visible due to long integration times (60\,s per point). We used a fitting function to estimate the visibilities and their associated asymptotic standard errors. We obtained net (raw) visibilities of $99\pm2\%$ ($83\pm1\%$). Here the corresponding subtracted noise, or accidental coincidence events, is only due to intrinsic dark counts in the employed APDs.}
\end{figure}

Moreover, when fibers are used as transmission channels, the paired photons experience a relative phase accumulated by the H and V modes along the propagation arising from birefringence. In the case where the photons are separated and arrive each at one user's place, the entangled state reads: $\ket{\Psi}= \frac{1}{\sqrt{2}}[\ket{H_a,V_b} + e^{i(\phi_a+\phi_b)}\ket{V_a,H_b}]$, where $\phi_j$ corresponds to the phase between $\ket{H_j}$ and $\ket{V_j}$ accumulated over the $j$ channel. Such a state is maximally entangled. However the presence of the phase, when not canceled, implies using appropriate settings for Alice and Bob's analyzers instead of those mentioned above. Such an unknown setting configuration can be avoided by placing a Soleil-Babinet (SB) compensator in one of the channels, say in Alice's as depicted in \figurename{~\ref{fig:montage}}. The SB thickness through which Alice's photons travel is chosen so as to perfectly balance the accumulated phase, \textit{i.e.}, $\phi_a + \phi_b + \phi_{SB}=n \pi$. This condition can be easily obtained when both users chose the $\{D,A\}$ basis: Alice only hat to adjust the SB thickness towards maximizing or minimizing the coincidence rate between her and Bob's detectors.

In \figurename{~\ref{fig:Bells}} we show the obtained four raw coincidence rates in the $\{H,V\}$ ($\theta_{HWP_a}= 0$ and 45$^\circ$) and $\{D,A\}$ ($\theta_{HWP_a}= 22.5$ and 67.5$^\circ$) basis, respectively. Using an appropriate fitting function, we estimated the raw visibilities to be of $83\pm1\%$ for the four interference patterns. The corresponding net visibilities, after subtraction of the accidental coincidences, reach 99$\pm$2\%. The latter values are among the best ever reported for similar configurations, which proves the relevance of our overall approach.
Note that the accidentals are initiated only by the dark counts in the employed detectors. However, when true applications have to be considered, only raw visibilities matter since background noise cannot be subtracted. In our experiment, this limitation is only due to poor detection efficiencies and high dark count probabilities, and using better detectors is the only solution. In this context, superconducting single photon counting modules~\cite{Verevkin_ultrafast_2004,Engel_Supra_2004}, up-conversion based detectors for which the pump wavelength is higher than the signal to avoid any pump induced background noise~\cite{langrock_up_2005,tanzilli_interface_2005,tournier_up_2009}, or recently developed free-running~\cite{thew_free-running_2007} and rapid gating~\cite{zhang_practical_2009} methods for InGaAs APDs are possible inputs.

For a fair comparison with the sources discussed in the state of the art, we give in \tablename{~\ref{tab:inv}} the related conversion efficiencies and entanglement net visibilities~\cite{suhara_generation_2007,takesue_generation_2005,jiang_generation_2007,kawashima_type_I_2009,lee_generation_2006}.
\begin{table}[htb]
 \begin{tabular}{|l|l|l|l|}
 \hline
 Ref. & Configuration & Conv. Efficiency & Net Visibilities\\
 \hline
 H. Takesue {\it et al.}~\cite{takesue_generation_2005}& 1 type-I PPLN/W & $3.6 \times 10^{-6}$ & 82\%\\
 \hline
 Y.K. Jiang {\it et al.}~\cite{jiang_generation_2007} & 2 type-I PPLN/W & $2 \times 10^{-9}$ $^\dagger$ & 93.5 $\pm$ 2.6\%\\
 \hline
 J. Kawashima {\it et al.}~\cite{kawashima_type_I_2009} & 1 type-I PPLN/W & $2.7 \times 10^{-7}$ & 83\%\\ 
 \hline
 K.F. Lee {\it et al.}~\cite{lee_generation_2006}& DSF & $3.2 \times 10^{-32}$ & 98.3\%\\
 \hline
  M. Medic {\it et al.}~\cite{medic_2010}& DSF & NA & 99.4 $\pm$ 1.2\%\\
 \hline
 T. Suhara {\it et al.}~\cite{suhara_generation_2007} & 1 type-II PPLN/W & $5.3 \times 10^{-10}$ & 90\%\\
 \hline
 A. Martin {\it et al.} & 1 type-II PPLN/W & 1.1 $\times$ 10$^{-9}$ & 99 $\pm$ 2\%\\
 \hline
\end{tabular} 
\caption{\label{tab:inv}Compared results obtained for various non-linear waveguide based sources. We give, for each considered configuration, the related internal conversion efficiency for the generators, the net visibility associated with a Bell inequality type test.\\$^\dagger$\footnotesize{For this configuration, which takes advantage of two cascaded interactions, namely second harmonic generation for preparing the pump field followed by SPDC along the same sample, the given value concerns the overall conversion efficiency. Therefore, this figure of merit is underestimated regarding only SPDC.}
}
\end{table}
As expected, we see from this table that our type-II source is less efficient than those based on type-I interactions. This has to do with the fact that we use a smaller non-linear coefficient ($d_{24}$) of lithium niobate compared to that used for type-I ($d_{33}$). However, the simple approach we have chosen enables a maximal exploitation of the available technological resources, leading to a near-perfect degree of entanglement, while maintaining a high brightness.

As a complementary information, finally note that the so-called Bell parameter $S$ can be extracted from our data to quantify the amount of violation of the Bell inequality~\cite{Bell_1964}. Following the reformulation of the Bell inequality proposed by Clauser, Horne, Shimony, and Holt (CHSH)~\cite{clauser_proposed_1969}, and thanks to the fits of the four interference patterns (see \figurename{~\ref{fig:Bells}}), we obtain $S_{exp}=2.80\pm0.03$, meaning that the BCHSH inequality is violated by more than 25 standard deviations.

\section{Perspectives}

This source can be upgraded by using a fiber birefringence delay compensator, to avoid using our bulk device and further reduce the losses in an even more compact setup. For that we can either take advantage of a fibered PBS and two fibered Faraday mirrors to replace the ``QWP + standard mirror'' sets, or of a high birefringence polarization maintaining fiber. In both cases, we can adjust the delay between the two polarization modes by applying a mechanical constraint on the fibers, or by tuning the temperature.

Such a source is obviously fully extendable to the 1550 nm telecom window that corresponds to the minimum absorption of standard single mode fibers. To reach this emission wavelength, one has to change both the poling period of the PPLN substrate and the pump wavelength accordingly. For a pump wavelength of about 780 nm and for similar operation temperatures, we have calculated the PPLN grating step to be on the order of 9.1\,$\mu$m. As a consequence for the detection, the Ge-APDs, which is essentially insensible to the 1550 nm wavelength, has to be replaced by the above mentioned InGaAs-type APDs~\cite{thew_free-running_2007,zhang_practical_2009}. To take advantage of a passive quenching mode and of higher maximum counting rates, one may employ superconducting~\cite{Verevkin_ultrafast_2004,Engel_Supra_2004} or up-conversion detectors~\cite{langrock_up_2005,tournier_up_2009,Thew_Tunable_SFG} instead.

Another possible improvement would be to make this source compatible with currently investigated quantum memories enabling true quantum networking applications. A relevant and very recently published review article can be found in Ref.~\cite{lvovsky_oqm_2009}. Nevertheless, it is worth noting that such quantum memories are both device and storage protocol dependent. We find in the literature, among others, solutions based on alkaline cold atom ensembles associated with electromagnetically induced transparency~\cite{chaneliere_rb_2005}, atomic vapors associated with off-resonant two-photon transitions~\cite{reim_qm_2010}, ion-doped crystals associated with atomic frequency combs~\cite{de-riedmatten_ss-qm_2008,chaneliere_tm_2009}, and ion-doped waveguides associated with photon-echoes~\cite{sinclair_qm-wg_2009}. These combinations offer various interaction wavelengths and linewidths, ranging from the visible to the telecom band, and from a few tens of MHz to almost 1 GHz, respectively. In any case, both the wavelength and bandwidth of the paired photons have to be adapted to the considered storage device and protocol. In particular, an additional filtering stage has to be employed so as to highly reduce the bandwidth by at least two orders of magnitude.
This can be done by using a Fabry-Perot cavity, provided the pump laser power is increased so as to recover an acceptable pair production rate within the limit of the considered experiment regarding the created mean number of pairs (should be $<$1) per relevant time window~\cite{halder_high_2008,pomarico_waveguide_2009}. This is where the advantage of integrated non-linear optics schemes lies in terms of efficiency. Since a few hundreds of mW should be enough, one has to pay extra attention to double pair emission within the time detection window that could lead to a decrease of the entanglement quality.

Finally, to complete the latter argument, Ti-in-diffused waveguides are appropriate candidates for addressing the issue of generating non-degenerate cross-polarization-entangled photon pairs. This could be made possible by an appropriate engineering of the QPM grating so as to satisfy two SPDC processes simultaneously, namely, down conversion of ordinary pump photon to either extraordinary signal and ordinary idler, or conversely, as proposed in Ref.~\cite{Thyagarajan_generation_2009}. Such a generator scheme should be of great interest in applications requiring, for instance, one of the entangled photons at an appropriate wavelength being used for local operation or for quantum storage in an atomic ensemble, and the other one at the typical wavelength of 1550\,nm for propagation through an optical fiber.

\section{Conclusion} 

We have demonstrated a narrow-band source of polarization entangled photon-pairs emitted at 1310\,nm based on a type-II PPLN waveguide and outlined possible improvements that could be provided with this approach. The obtained near-perfect quality of entanglement, which corresponds to the best value ever reported for guided-wave schemes, associated with high brightness and practicality make this source a promising element for long-distance quantum communication protocols, and highlights the high potential of non-linear integrated optics in this field. 

The experimental setup we have designed and employed enabled us to record more than a hundred and about 40 points for the HOM- and for the Bell-type measurements, respectively. Noticeably in the latter case, this indicates that the setup was free of any type of fluctuation (phase, intensity, \textit{etc.}) over the measurement time. This also proves both the very high control we have achieved on all the experimental parameters, and the relevance of our approach towards achieving a near perfect quality polarization entanglement waveguide-based source emitting at a telecom wavelength.

\ack

The authors acknowledge A. Thomas and M. P. De Micheli for fruitful discussions, and the European ERA-SPOT program ``WASPS'', l'Agence Nationale de la Recherche for the ``e-QUANET'' project ANR-09-BLAN-0333-01, the CNRS, le Conseil Regional PACA, and the University of Nice - Sophia Antipolis for financial supports.

\end{document}